\documentclass[final,5p,times,twocolumn,number,compress]{elsarticle}

\usepackage{amssymb}
\usepackage{lipsum}
\usepackage{amsmath,amssymb}
\usepackage{physics}
\usepackage{graphicx}
\usepackage{dcolumn}
\usepackage{bm}
\usepackage{hyperref}
\usepackage[caption=false]{subfig}
\usepackage{subcaption}
\usepackage{multirow}
\usepackage[dvipsnames]{xcolor}
\usepackage{color}
\usepackage{cancel}
\usepackage[normalem]{ulem}
\usepackage{array}
\usepackage[utf8]{inputenc}
\usepackage{graphicx} 
\usepackage[T1]{fontenc}
\usepackage{xr}
\usepackage{float}
\usepackage{placeins}
\usepackage{makecell}

\usepackage{natbib}
\usepackage{changes}

\usepackage{times}
\usepackage{microtype}
\usepackage{orcidlink}

\DeclareUnicodeCharacter{2212}{-}

\journal{Physics Letters B}

\begin{document}

\begin{frontmatter}

\title{Threshold-Aligned Pygmy Dipole Strength in Astrophysical $(n,\gamma)$ and $(\gamma,n)$ Reactions}

\author[first]{T. Ghosh}
\ead{tghosh@phy.hr}

\author[first]{A. Kaur}
\ead{akaur@phy.hr}

\author[first]{N. Paar}
\ead{npaar@phy.hr}


\affiliation[first]{organization={Department of Physics, Faculty of Science, University of Zagreb},
            addressline={Bijeni\v cka cesta 32}, 
            city={Zagreb},
            postcode={10000}, 
            country={Croatia}}  

\begin{abstract}

Reaction-rate calculations relevant to r-process nucleosynthesis depend sensitively on the nuclear $\gamma$-strength function ($\gamma$SF). Here we investigate the impact of low-lying pygmy dipole strength (PDS) in $(n,\gamma)$ and $(\gamma,n)$ reactions using $\gamma$SF based on relativistic nuclear energy density functional theory and propagate these strengths into Hauser--Feshbach statistical model calculations of the reaction rates. We show that considerable reaction-rate enhancements at temperatures relevant for r-process nucleosynthesis are governed by the alignment of the pygmy dipole strength energy with the neutron separation threshold $S_n$ rather than by the total low-energy strength. Consequently, nuclei such as $^{68}$Ni and $^{132}$Sn, where the PDS energy-$S_n$ alignment occurs, exhibit the strongest effects on reaction-rate enhancements. These results demonstrate that modeling reliable reaction rates in r-process nucleosynthesis necessitates accurate microscopic descriptions of low-energy dipole strength, in close synergy with experimental investigations in the vicinity of neutron threshold.

\end{abstract}

\end{frontmatter}
\section*{Introduction}
Radiative neutron capture (n,$\gamma$) and photoneutron ($\gamma$,n) reactions are of central importance in the fields ranging from medical applications \cite{Lima1998} and energy production \cite{Prelas2016} to astrophysical nucleosynthesis \cite{Burbidge1957}. In particular, they provide the fundamental pathways by which nearly all elements heavier than iron are synthesized through the $s$- and $r$-processes \cite{Burbidge1957,Arnould2007}. The impact on calculations of (n,$\gamma$) and ($\gamma$,n) reaction rates are profound, as the statistical Hauser–Feshbach framework \cite{Hauser1952} requires nuclear inputs, particularly the gamma-ray strength function ($\gamma$SF), nuclear level density (NLD) and optical model potentials (OMP). The uncertainties resulting from these inputs directly propagate into nucleosynthesis calculations, making the inclusion of consistent and accurate description of these inputs essential for reliable astrophysical modeling. Uncertainties in the $\gamma$SF and NLD dominate the error budget of calculated neutron-capture rates, while those from the OMP are comparatively smaller \cite{Liddick2016, Sangeeta2022}. The $\gamma$SF, which describes the average nuclear response to $\gamma$-ray absorption and emission, together with the NLD, governs the probability of photon decay or capture. For exotic nuclei, the $\gamma$SF is usually extrapolated from stable systems using phenomenological Lorentzian descriptions of the giant dipole resonance (GDR) \cite{Kopecky1990}. However, significant deviations from this standard behavior have been observed in both stable and unstable nuclei. In particular, enhancements near the neutron separation energy \cite{Savran2013} and at very low transition energies \cite{Voinov2004} have been reported, with potentially dramatic consequences for predicted (n,$\gamma$) capture rates \cite{GORIELY199810,Larsen2010,Goriely2004}. 
These sensitivities highlight the need for a detailed understanding of dipole strength distributions, particularly at low excitation energies, where modifications to the $\gamma$SF can strongly influence reaction rates.

The study of low-lying dipole excitations in nuclei has attracted considerable interest over the past two decades, both experimentally and theoretically, because of their profound connection to nuclear structure and astrophysical processes \cite{GORIELY199810,Paar_2007,PhysRevC.85.041304,SAVRAN2013210,LANZA2023104006}. In particular, the pygmy dipole strength (PDS) represents an accumulation of the electric dipole (E1) strength at excitation energies below the giant dipole resonance (GDR). This additional strength is especially prominent in neutron-rich nuclei and its existence is also predicted in nuclei close to the proton drip lines \cite{Paar2005}, but rather complex structure of PDS is still under discussion \cite{PhysRevLett.103.032502,PhysRevLett.125.102503,PhysRevC.89.041601,LANZA2023104006,refId0}.
The presence of PDS is correlated not only with the neutron excess in finite nuclei, but also with fundamental properties of nuclear matter, such as neutron skin thickness and dipole polarizability, thus providing constraints on the nuclear symmetry energy and its slope parameter, which are crucial inputs for the nuclear equation of state \cite{Piekarewicz2006,PhysRevC.76.051603,PhysRevC.81.041301,PhysRevC.84.027301,Tamii2011}. Beyond its nuclear structure implications, the PDS plays an essential role in nuclear astrophysics. Since the $r$-process proceeds through very neutron-rich nuclei far from stability, where experimental constraints on dipole strength are scarce, theoretical predictions of PDS properties directly feed into astrophysical simulations. The enhanced low-energy $\gamma$-ray strength associated with PDS can significantly impact radiative neutron capture rates relevant for rapid neutron capture ($r$-process) and intermediate neutron capture ($i$-process) nucleosynthesis \cite{Daoutidis2012}.

In this letter, we investigate the role of the PDS both in (n,$\gamma$) and ($\gamma$,n) reactions by considering $\gamma$SFs in the framework of relativistic nuclear energy density functional. In particular, by systematically analyzing dipole transitions along the isotopic chains {$^{56-82}$Ni and $^{100-150}$Sn}, we quantify the enhancement of PDS, evaluate its contribution to low-energy $\gamma$-ray strength functions, and discuss the resulting impact on 
(n,$\gamma$) and ($\gamma$,n) reaction rates relevant in astrophysical environments. The aim is also to investigate how the position of the PDS excitation energy with respect to the neutron threshold impacts reaction-rate calculations.

\section*{Methodology}

Theoretical investigations of electric dipole response in nuclei, including PDS, have been carried out in the past in a variety of frameworks (for an extensive review see Ref. \cite{LANZA2023104006}), including quasiparticle random phase approximation (QRPA) based on non-relativistic \cite{LANZA2023104006,PhysRevC.81.041301,Liu_2025} or relativistic \cite{Paar_2007,PhysRevC.67.034312} energy density functionals (EDF), various beyond (Q)RPA methods including complex configurations \cite{PhysRevLett.105.022502,TERTYCHNY2007159,PhysRevC.79.054312,Tsoneva_2016,TSONEVA2025123114,Leoni_2019,SAVRAN201816,PhysRevC.90.014310},
time-dependent Hartree-Fock-Bogoliubov theory \cite{PhysRevC.82.034306}, configuration interaction shell model \cite{sieja2017low,PhysRevC.85.051601}, microscopic transport model \cite{PhysRevC.88.044610}, interacting boson model \cite{PhysRevC.85.064315}, etc. These studies have consistently established the existence of low lying E1 strength in medium and heavy mass nuclei, with its magnitude growing with increasing neutron to proton asymmetry. Recently developed finite-temperature relativistic quasiparticle random phase approximation (FT-RQRPA) \cite{PhysRevC.109.014314,3t7h-nds7,96g9-1ff5} provides a powerful self-consistent framework to investigate also the temperature effects on PDS and collective nuclear excitations.

In this work, we employ the zero-temperature limit of FT-RQRPA \cite{PhysRevC.109.014314}, derived from the relativistic nuclear energy density functional with point-coupling interaction \cite{Niksic_2008}. Within RQRPA, nuclear excited states $|\nu\rangle$ are described as small-amplitude oscillations
around the ground state $|0\rangle$, obtained using the relativistic Hartree--Bogoliubov model~\cite{NIKSIC20141808,Niksic_2008,PhysRevC.67.034312}.
The RQRPA eigenvalue problem reads
\begin{equation}
\begin{pmatrix}
A & B \\
-B^{*} & -A^{*}
\end{pmatrix}
\begin{pmatrix}
X^{\nu} \\[1mm] Y^{\nu}
\end{pmatrix}
=
E_{\nu}
\begin{pmatrix}
1 & 0 \\ 0 & -1
\end{pmatrix}
\begin{pmatrix}
X^{\nu} \\[1mm] Y^{\nu}
\end{pmatrix},
\end{equation}
where $E_{\nu}$ denotes the excitation energy of the $\nu$-th mode, and 
$X^{\nu}$ and $Y^{\nu}$ are the forward and backward amplitudes.
The matrices $A$ and $B$ encode the residual two-body interaction, derived self-consistently from the underlying energy-density functional \cite{PhysRevC.67.034312,PhysRevC.109.014314}.


By using the isovector electric dipole operator \cite{Paar_2007}, 
%
%
the discrete dipole transition strength function is obtained,
\begin{equation}
S(E1;E) = \sum_{\nu} S(E1;0\!\rightarrow\!\nu)\, 
\delta(E - E_{\nu}).
\end{equation}
To account for spreading effects observed in experimental dipole strengths, the discrete RQRPA strengths are folded with a Lorentzian function \cite{Daoutidis2012,Drozdz1990,96g9-1ff5}, yielding the continuous response
\begin{equation}
R(E1;E)
=
\sum_{\nu}
\frac{1}{2\pi}
\frac{\Gamma(E)\,
S(E1;0\!\rightarrow\!\nu)}
     {\left(E - E_{\nu} - \Delta(E)\right)^{2} + \Gamma(E)^{2}/4},
\label{eq:cont_response}
\end{equation}
where $\Gamma(E)$ and $\Delta(E)$ denote the energy-dependent width and the corresponding energy shift, respectively, which are given by the imaginary and real parts of the self-energy \cite{Drozdz1990}. 
%
%
The gamma strength function is given as
\begin{equation}
f_{E1}(E)
=
\frac{16\pi e^{2}}{27\,\hbar^{3} c^{3}}
\, R(E1;E),
\end{equation}
where $e$ is the elementary charge. 

The width $\Gamma(E)$ can be obtained from measured particle and hole decay widths as \cite{Drozdz1990},
\begin{equation}
\Gamma(E) 
= \frac{1}{E} 
  \int_{0}^{E} d\epsilon\,
  [\,\gamma_{p}(\epsilon) 
   + \gamma_{h}(\epsilon - E)\,]
  (1 + C_{G}),
\end{equation}
and the energy shift follows from the dispersion relation
\begin{equation}
\Delta(E)
= \frac{1}{2\pi}\,
  \mathcal{P}
  \int_{-\infty}^{\infty}
  dE'\,
  \frac{\Gamma(E')} {E' - E}\,
  (1 + C_{E}),
\end{equation}
with interference coefficients $C_{G}$ and $C_{E}$, which are constrained by minimizing the $\chi^2$ objective function to reproduce experimental photoabsorption and photoneutron $\gamma$SF \cite{goriely2019reference}. The values of $C_{G}$ and $C_{E}$ are $-0.584$ and $-1.149$, respectively, as obtained in recent RQRPA calculation in Ref. \cite{96g9-1ff5}.


\begin{figure*}[htbp]
    \centering
    \includegraphics[width=0.99\textwidth]{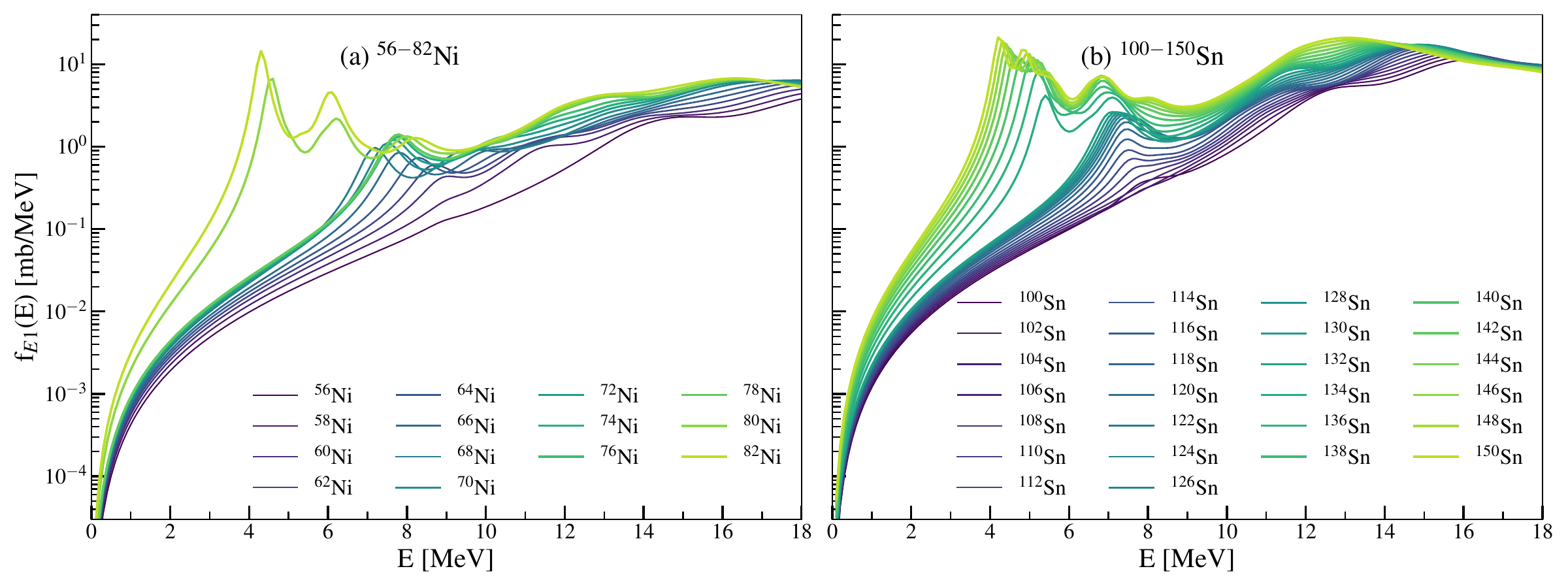}
\caption{Electric dipole $\gamma$-ray strength functions $f_{E1}(E)$ for even–even $^{56-82}$Ni (a) and $^{100-150}$Sn (b) isotopic chains calculated with the DD-PCX energy density functional.}
\label{fig:gsf}
\end{figure*}

Using the isovector dipole response from the folded RQRPA strength function given in Eq. (\ref{eq:cont_response}), one can consistently calculate radiative neutron-capture and photoneutron cross sections within the statistical Hauser–Feshbach model~\cite{Hauser1952}. 
In this framework, the respective cross section is governed by the $\gamma$-ray strength function together with the nuclear level density and the standard kinematic and transmission ingredients. Although additional multipoles can contribute, $\gamma$ decay is typically dominated by E1 transitions, making the RQRPA-derived dipole strength a key microscopic input to the cross-section calculation. 
For astrophysical applications, the observable of interest is Maxwellian-averaged value of the energy-dependent cross section (MACS), defined at stellar temperature $T$ as
\begin{equation}\label{eq:macs}
\langle \sigma v \rangle_T 
= 
\left( \frac{8}{\pi m} \right)^{1/2}
\frac{1}{(kT)^{3/2}}
\int_{0}^{\infty}
\sigma(E)\,E\,e^{-E/kT}\, dE ,
\end{equation}
where $m$ is the reduced mass of the neutron–target nucleus system and $k$ the Boltzmann constant.  The MACS values derived from this expression serve as essential inputs to $r$-process network calculations and nucleosynthesis modeling~\cite{Iliadis_book}.

In the present study, first the $\gamma$SFs are calculated from the RQRPA dipole transition strength, based on density-dependent point coupling interaction DD-PCX~\cite{PhysRevC.99.034318}, and then implemented in the reaction code \textsc{TALYS}-2.0~\cite{koning_talys}, which is subsequently used to compute neutron-capture and photoneutron cross sections and the corresponding Maxwellian-averaged reaction rates. All other nuclear-structure ingredients required by \textsc{TALYS}, including nuclear masses, optical-model potentials (OMP), and nuclear level densities (NLD), were taken from the default parameter sets of the code ~\cite{koning_talys}. This setup allows us to explore the impact of PDS on both neutron-capture and photoneutron cross sections and astrophysical reaction rates by performing nuclear reaction calculations with two different $\gamma$SF inputs: one including only the giant dipole resonance (GDR), and another one containing GDR and also the pygmy dipole strength (GDR+PDS).

\section*{Results and Discussions}

Figure~\ref{fig:gsf} shows the calculated E1 $\gamma$-ray strength functions for even-even $^{56-82}$Ni and $^{100-150}$Sn isotopes using the RQRPA with DD-PCX functional, including both the pygmy dipole strength and GDR. With increasing neutron number, both isotopic chains exhibit a gradual lowering of the GDR centroid and a systematic buildup of low-energy dipole strength around neutron separation energy. This enhancement is moderate in Ni isotopes but becomes pronounced in neutron-rich Sn isotopes. This evolution of the $\gamma$SF reflects the increasing softness of the dipole mode and the overall magnitude and trend of the $\gamma$SFs demonstrate the sensitivity of dipole response to isovector properties of the functional and highlight the increasing role of low-energy E1 strength for nuclei approaching the neutron drip line, which has direct implications for $(n,\gamma)$ and $(\gamma,n)$ reaction rates in r-process stellar environments.

\begin{figure*}[htbp]
    \centering
    \includegraphics[width=0.8\textwidth]{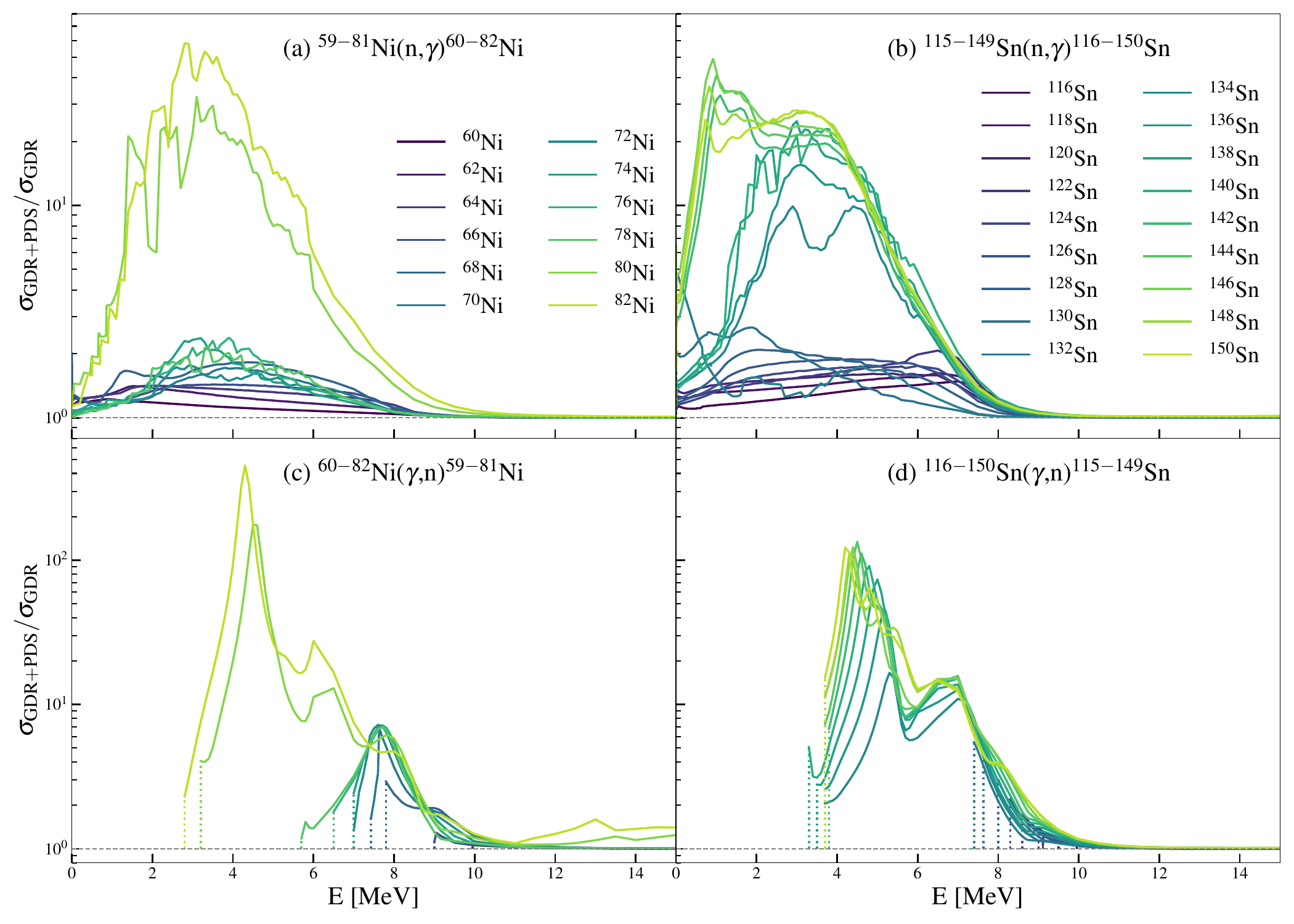}
\caption{
Energy-dependent ratios of the cross-sections calculated with and without pygmy dipole strength in $(n,\gamma)$ (a), (b) and $(\gamma,n)$ (c), (d) reactions
for even–even $^{60\text{–}82}$Ni and $^{116\text{–}150}$Sn compound nuclei.}
\label{fig:e_cs_ratio}
\end{figure*}

In the following we investigate in more details the role of PDS in the reaction rates.
Figure \ref{fig:e_cs_ratio} presents the energy-dependent ratios of 
neutron-capture $(n,\gamma)$ and photoneutron $(\gamma,n)$ reaction
cross sections along the Ni and Sn isotopic chains,
obtained using $\gamma$SFs including both GDR and PDS contributions with respect to those only with GDR, $\sigma$(GDR+PDS)/$\sigma$(GDR).  The upper two panels show the $(n,\gamma)$ ratios for (a) Ni isotopes and (b) Sn isotopes, while the lower two panels display the corresponding $(\gamma,n)$ ratios for (c) Ni and (d) Sn isotopes. For the photoneutron case, the ratios are shown only above the neutron separation energy, since the $(\gamma,n)$ channel opens at this threshold.
For all isotopes considered, the inclusion of the PDS produces a systematic enhancement of the reaction cross section ratios above unity over a finite low-energy interval. The magnitude of the enhancement is strongly isotope dependent, with heavier and more neutron-rich isotopes exhibiting significantly larger ratios and slightly broader enhancement regions. In several neutron-rich nuclei, the cross section ratio exceeds values of 50--70 for $(n,\gamma)$ reactions and 200--400 for $(\gamma,n)$ reactions in the vicinity of the PDS region, indicating a substantial impact of low-energy E1 strength on reaction dynamics. At higher energies, the ratios gradually approach unity, reflecting the dominance of higher-energy components of the dipole strength where the relative contribution of PDS is reduced. These trends demonstrate that the presence of PDS modifies both the magnitude and the energy dependence of reaction cross sections in a non-uniform manner across isotopic chains.

\begin{figure*}[htbp]
    \centering
    \includegraphics[width=0.8\textwidth]{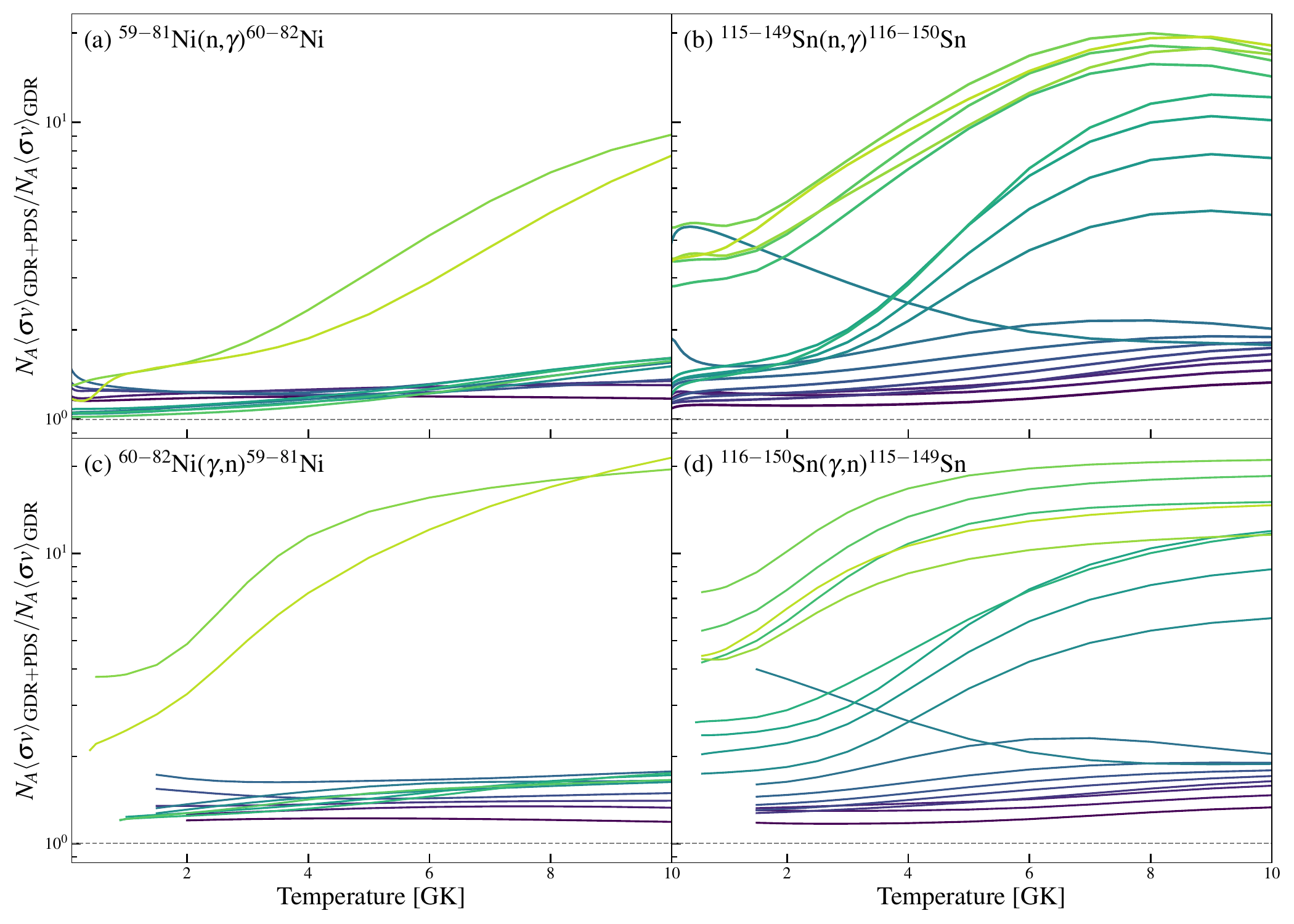}
\caption{
Temperature-dependent reaction-rate ratios $N_A \langle \sigma v \rangle_{\mathrm{GDR+PDS}} / N_A \langle \sigma v \rangle_{\mathrm{GDR}}$ for $(n,\gamma)$  (a), (b) and $(\gamma,n)$ (c), (d) reactions of even–even $^{60\text{–}82}$Ni and $^{116\text{–}150}$Sn. The same color coding like in Fig.~\ref{fig:e_cs_ratio} is used.}
\label{fig:e_astro_ratio}
\end{figure*}

\begin{figure*}[htbp]
    \centering
    \includegraphics[width=0.78\textwidth]{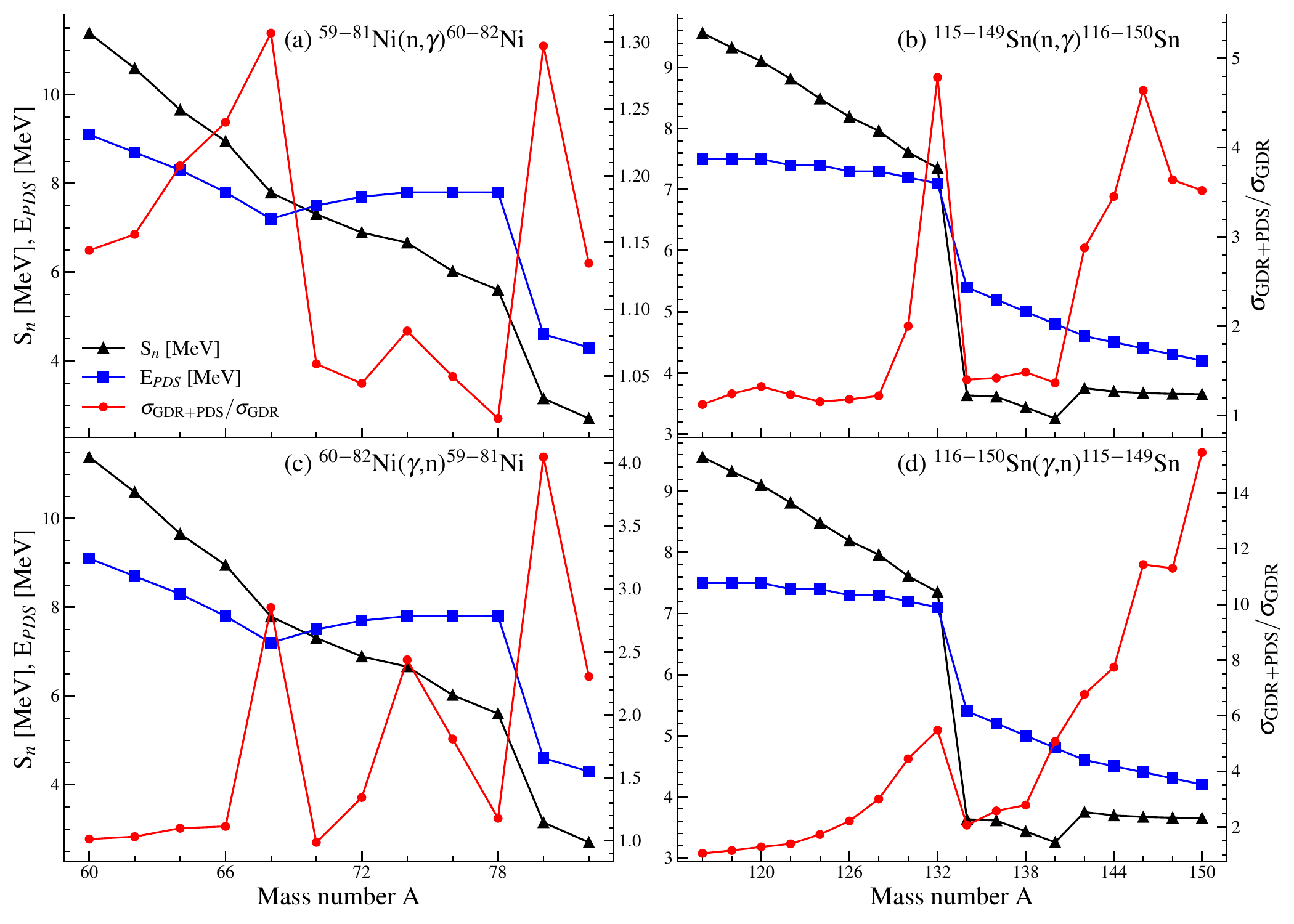}
\caption{
Neutron separation energies ($S_n$) and PDS peak energies ($E_{\mathrm{PDS}}$) for Ni and Sn isotopes (left $y$-axis).
Ratios of $(n,\gamma)$ (a), (b) and $(\gamma,n)$ (c), (d) cross sections calculated with and without pygmy contributions (right $y$-axis), evaluated at 30\,keV and at $S_n + 30$\,keV, respectively, for the compound nuclei $^{60\text{–}82}$Ni and $^{116\text{–}150}$Sn.} 

\label{fig:ratio_cs_ng_gn}
\end{figure*}

A similar behavior is observed for the astrophysical reaction rates calculated using Eq.(\ref{eq:macs}). Figure~\ref{fig:e_astro_ratio} shows the ratios of temperature dependent rates $N_A \langle \sigma v \rangle_{\mathrm{GDR+PDS}} / N_A \langle \sigma v \rangle_{\mathrm{GDR}}$, obtained with and without inclusion of PDS strength, for $(n,\gamma)$ and $(\gamma,n)$ reactions. The temperature range up to $T = 10$~GK is considered in order to cover the astrophysical conditions relevant for neutron-capture and photodisintegration processes.
 For both Ni and Sn isotopes, the inclusion of PDS leads to an increase in reaction rates over a wide temperature interval, with the effect becoming increasingly significant toward neutron-rich nuclei. Although cross-section ratios can reach values of 50--400 as shown in Fig. \ref{fig:e_cs_ratio}, thermal averaging over a broad energy interval reduces the net rate enhancement to the rate ratios factors of 1--20. 
For $^{68}$Ni and $^{132}$Sn, both the reaction rate and the cross-section ratio start at comparatively higher values than those of their neighboring nuclei. This behavior can be attributed to the fact that, in these two nuclei, the peak of the pygmy dipole strength closely coincides with the neutron separation energy $S_n$. In the $(n,\gamma)$ reaction, the compound nucleus is formed at energies close to the neutron separation energy (plus the small excitation energy at low incident neutron energies $E_{exc}$). Similarly, for the $(\gamma,n)$ reaction, the reaction threshold is determined by $S_n$. Therefore, in both cases, particularly at low temperatures where the reaction rate is dominated by very low-energy cross-sections, the relevant $\gamma$-ray transitions involve energies close to $S_n$ + $E_{exc}$. When $S_n$ closely coincides with the PDS peak energy ($E_{PDS}$), the impact of the pygmy strength becomes most pronounced. This effect is more pronounced in $^{132}$Sn due to its stronger and more concentrated PDS around the neutron separation energy, compared to $^{68}$Ni. As the temperature increases, the Maxwell-Boltzmann distribution broadens and becomes less localized around very low energies. Consequently, a wider energy region contributes to the reaction rate. In this regime, nuclei with larger overall pygmy strength exhibit enhanced ratios, irrespective of the exact position of the pygmy peak. This behavior becomes increasingly significant in more neutron-rich nuclei, where the pygmy contribution is comparably stronger.
Owing to the alignment of $E_{\mathrm{PDS}}$ and $S_n$, $^{68}$Ni and $^{132}$Sn exhibit comparatively large ratios at low temperatures. With increasing temperature, however, the ratios reduce, showing a trend distinct from that of the other isotopes. This behavior is particularly pronounced in $^{132}$Sn.
Overall, these results demonstrate that PDS plays an important role in shaping both energy-dependent cross sections and temperature-dependent reaction rates, and that its accurate description is essential for reliable modeling of nuclear reactions and nucleosynthesis in neutron-rich environments.

\begin{figure*}[htbp]
    \centering
    \includegraphics[width=0.78\textwidth]{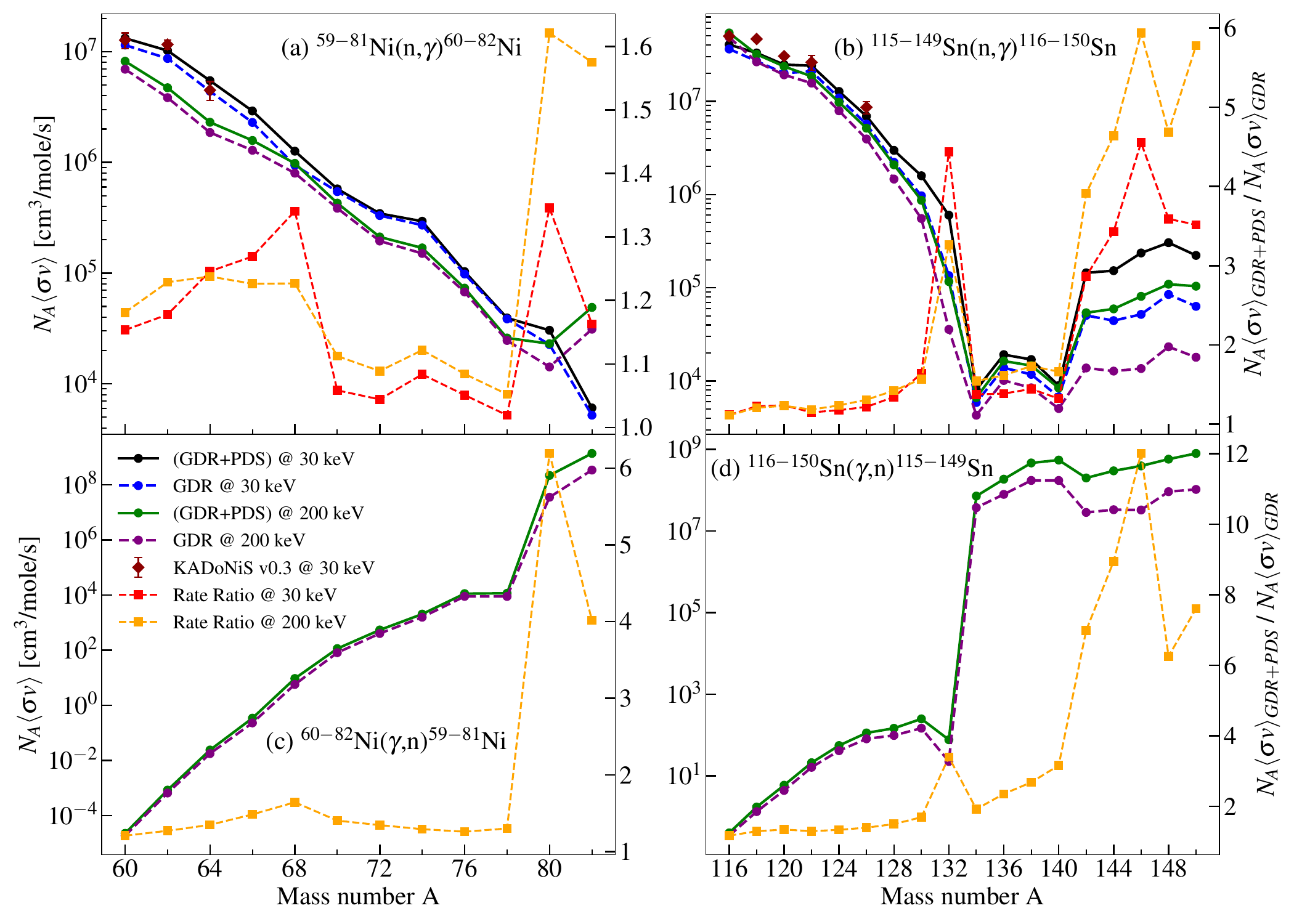}
\caption{
The rates for $(n,\gamma)$ (a), (b) and $(\gamma,n)$ (c), (d) reactions for Ni and Sn istopic chains (left $y$-axis).  
Ratios of reaction rates calculated with and 
without pygmy dipole strength, $N_A\langle\sigma v\rangle_{\mathrm{GDR+PDS}}/N_A\langle\sigma v\rangle_{\mathrm{GDR}}$, are shown for even–even $^{60\text{–}82}$Ni and $^{116\text{–}150}$Sn compound nuclei (right $y$-axis). The $(n,\gamma)$ rates are evaluated at $kT=30$ and 200\,keV, while the $(\gamma,n)$ rates are shown at $kT=200$\,keV. Comparison with available KADoNiS v0.3 \cite{kadonis} data at $kT=30$ keV is included for the $(n,\gamma)$ reaction rates.
}
\label{fig:ratio_astro_ng_gn}
\end{figure*}

To illustrate the role of neutron threshold–aligned pygmy strength, Fig.~\ref{fig:ratio_cs_ng_gn} shows the PDS peak excitation energy calculated within the RQRPA, together with the neutron separation energy $S_n$, for the Ni and Sn isotopic chains (left $y$-axes). The plotted values of $S_n$ are obtained from experimental masses, or from theoretical calculations in the framework of self-consistent Hartree-Bogoliubov theory (RHB) and RQRPA approximation when experimental data are unavailable.
The corresponding ratios of neutron-capture $(n,\gamma)$ and photoneutron $(\gamma,n)$ cross sections calculated with and without inclusion of pygmy dipole strength, $\sigma_{\mathrm{GDR+PDS}}/\sigma_{\mathrm{GDR}}$, are shown on the right $y$-axes. The upper two panels, (a) and (b), present results for the Ni and Sn chains, respectively, for the $(n,\gamma)$ reaction evaluated at $E_n=30$~keV, while the lower two panels, (c) and (d), show the corresponding $(\gamma,n)$ ratios evaluated at $E_\gamma = S_n + 30$~keV. 
As the neutron number increases along both isotopic chains, $S_n$ decreases and the PDS peak shifts systematically toward lower energies, with $S_n$ and PDR energies crossing around the (sub)shell closure. Pronounced enhancements of the cross-section ratios are observed for $^{68}$Ni and $^{132}$Sn, reflecting an alignment between $E_{\mathrm{PDS}}$ and $S_n$. 
For the neighboring isotopes, however, the ratios are significantly smaller because the PDS strength is shifted away from the neutron threshold, reducing its impact on the $(n,\gamma)$ cross section at $30$ keV and on the $(\gamma,n)$ channel at $S_n + 30$ keV.
The role of PDS strength increases further for more neutron-rich nuclei, where the neutron separation energy $S_n$ lowers and the pygmy dipole strength becomes more pronounced and spatially extended (see also Fig. \ref{fig:gsf}). In this regime, a significant fraction of the low-energy $E1$ response develops near the neutron threshold, making both $(n,\gamma)$ and $(\gamma,n)$ channels increasingly sensitive to its detailed structure. The pygmy mode therefore enhances the reaction strength in the astrophysically relevant threshold region.

Figure~\ref{fig:ratio_astro_ng_gn} presents the corresponding ratios of astrophysical reaction rates calculated with and without pygmy dipole strength for the same Ni and Sn isotopes. Panels (a) and (b) show the $(n,\gamma)$ rates (left $y$-axes) at a thermal energy of $kT=30$ and 200 keV, representative of astrophysically relevant temperatures for neutron-capture processes, while panels (c) and (d) display the $(\gamma,n)$ rates at $kT=200$~keV. For the $(\gamma,n)$ case only 200 keV temperature is considered because, at lower temperatures such as 30 keV, the relevant low-energy contributions are energetically inaccessible, whereas at 200 keV the tail of the Maxwell–Boltzmann distribution allows a significant contribution from higher-energy states. Comparison with KADoNiS v0.3 \cite{kadonis} data at 30 keV for the $(n,\gamma)$ rates further validates the reliability of the present calculation. Figure~\ref{fig:ratio_astro_ng_gn} also shows
ratios of reaction rates calculated with and without pygmy dipole strength, $N_A\langle\sigma v\rangle_{\mathrm{GDR+PDS}}/N_A\langle\sigma v\rangle_{\mathrm{GDR}}$ (right $y$-axes).
Inclusion of pygmy strength leads to a systematic enhancement of $(n,\gamma)$ rates, particularly for neutron-rich nuclei $^{68}$Ni and $^{132}$Sn, where the compound-nucleus excitation energy $E_{\mathrm{exc}}\approx S_n+30$~keV overlaps with the PDS region. Although microscopic cross-section ratios at the pygmy peak can reach values of several tens, thermal averaging over the Maxwell--Boltzmann distribution reduces the net rate enhancement to about 1.5 in Ni and up to 4--5 in Sn. For the $(\gamma,n)$ reaction, only temperatures around $kT=200$~keV yield non-negligible contributions, since neutron emission requires photon energies above $S_n$. As $S_n$ decreases toward neutron-rich nuclei with $S_n \approx 2$--$4$~MeV (see also figs. \ref{fig:e_astro_ratio}, \ref{fig:ratio_cs_ng_gn}), the overlap between the high-energy tail of the thermal distribution and the pygmy region increases, producing rapidly rising rate ratios, especially in the Sn isotopes. These results demonstrate that low-lying dipole strength significantly influences both neutron-capture and photoneutron reaction rates under $r$-process conditions and must be properly accounted for in reliable nucleosynthesis modeling.

\section*{Conclusions}

In this work we have demonstrated, through a consistent microscopic analysis of dipole strength functions, cross sections, and astrophysical reaction rates, that the low-lying PDS plays a quantitatively significant and strongly localized role in radiative neutron-capture $(n,\gamma)$ and photodisintegration $(\gamma,n)$ reactions relevant for the $r$-process. Presented calculations, based on description of $\gamma$SFs in the relativistic EDF framework with density dependent point coupling interaction, show that the impact of PDS is not only governed by its total strength but also by its energy alignment with neutron separation energy $S_n$: nuclei such as $^{68}$Ni and $^{132}$Sn, where the pygmy peak lies closest to the neutron-emission threshold, exhibit the largest enhancement of cross sections and reaction rates, while neighboring isotopes with larger pygmy -- neutron threshold energy gap show much weaker sensitivity.
The results also reveal that as nuclei become more neutron-rich and their neutron separation energies $S_n$ fall in the 2–4 MeV range, the pygmy strength becomes more pronounced and extends into the energy region where both $(n,\gamma)$ and $(\gamma,n)$ channels are particularly sensitive to the detailed structure of the $E1$ response.

Temperature effects further clarify this mechanism. At $T=30$\,keV, neutron captures probe $\gamma$SF at excitation energies $E_{\rm exc} \simeq S_n + kT$, making sensitivity to the PDS contingent on whether its centroid overlaps this narrow energy window. Even though the local cross-section enhancement at the pygmy peak can reach factors of 20–30, thermal averaging suppresses the rate modification to $\sim 1.5$ for Ni and up to $4$–$5$ for Sn. In contrast, $(\gamma,n)$ reactions involve energies of the order $S_n$ itself and therefore depend on the high-energy tail of the Maxwell–Boltzmann distribution. At such low temperatures, the nuclei with very small $S_n$ acquire any measurable PDS contribution, a feature that becomes more visible at higher thermal energies such as $T=200$\,keV, where the tail begins to access the pygmy region. These results collectively demonstrate that the influence of PDS on $r$-process reaction flows is governed simultaneously by nuclear-structure evolution along isotopic chains and by the thermal physics of the astrophysical environment.

Overall, this study highlights that reliable predictions of $(n,\gamma)$ and $(\gamma,n)$ reactions relevant for $r$-process abundances require models capable of resolving detailed properties the fine structure of pygmy dipole strength near threshold—not merely the gross GDR properties. Because the pygmy mode becomes increasingly prominent precisely where $S_n$ decreases most rapidly along the $r$-process path, its effect on reaction rates is both localized and decisive. The sensitivity patterns identified here underscore the need for
further systematic microscopic studies based on advanced theoretical approaches including assessment of theoretical uncertaintes in PDS, and novel experimental constraints targeting low-energy dipole strength in neutron-rich nuclei. As $r$-process nucleosynthesis paths continue to move toward the drip line with upcoming radioactive-beam facilities, the accurate treatment of pygmy strength will remain essential for predicting the formation of the heavy elements in the cosmos.

This work is supported by the Croatian Science Foundation under the project Relativistic Nuclear Many-Body Theory in the Multimessenger Observation Era (HRZZ-IP-2022-10-7773). N.P. acknowledges support by the project “Implementation of cutting-edge research and its application as part of the Scientific Center of Excellence for Quantum and Complex Systems, and Representations of Lie Algebras“, Grant No. PK.1.1.10.0004, co-financed by the European Union through the European Regional Development Fund - Competitiveness and Cohesion Programme 2021-2027.

\bibliographystyle{elsarticle-num}

\bibliography{bibliography}

\end{document}